\documentclass[12pt]{article}
\usepackage{graphicx}
\usepackage{hyperref}
\usepackage{authblk} 
\usepackage[top=2cm,bottom=2.5cm,left=2cm,right=2cm]{geometry}

\begin{document}
\title{Organisation of signal flow in directed networks}
\author[1,4]{M. B\'anyai} 
\author[2]{L. N\'egyessy} 
\author[1,3,*]{F. Bazs\'o} 
\affil[1]{\small{KFKI Research Institute for Particle and Nuclear Physics of the Hungarian Academy of Sciences, H-1525 Budapest, P.O. Box 49., Hungary}}
\affil[2]{\small{Neurobionics Research Group, Hungarian Academy of Sciences - P\'eter P\'azm\'any Catholic University - Semmelweis University, T\H{u}zolt\'o u. 58, H-1094 Budapest, Hungary}}
\affil[3]{\small{SU-Tech College of Applied Sciences, Subotica, Marka Ore\v{s}kovi\'ca 16, 24000 Subotica, Serbia}}
\affil[4]{\small{Budapest University of Technology and Economics, Faculty of Electrical Engineering and Informatics, Department of Measurement and Information Systems, Budapest, Hungary}}
\affil[*]{\texttt{e-mail: bazso@mail.kfki.hu}}
\date{ }
\maketitle

\begin{abstract}
Confining an answer to the question whether and how the coherent operation of network elements is determined by the the network structure is the topic of our work. We map the structure of signal flow in directed networks by analysing the degree of edge convergence and the overlap between the in- and output sets of an edge. Definitions of convergence degree and overlap are based on the shortest paths, thus they encapsulate global network properties. Using the defining notions of convergence degree and overlapping set we clarify the meaning of network causality and demonstrate the crucial role of chordless circles. In real-world networks the flow representation distinguishes nodes according to their signal transmitting, processing and control properties. The analysis of real-world networks in terms of flow representation was in accordance with the known functional properties of the network nodes. It is shown that nodes with different signal processing, transmitting and control properties are randomly connected at the global scale, while local connectivity patterns depart from randomness. Grouping network nodes according to their signal flow properties was unrelated to the network's community structure. We present evidence that signal flow properties of small-world-like, real-world networks can not be reconstructed by algorithms used to generate small-world networks. Convergence degree values were calculated for regular oriented trees, and its probability density function for networks grown with the preferential attachment mechanism. For Erd\H{o}s-R\'enyi graphs we calculated both the probability density function of convergence degrees and of overlaps. 
\end{abstract}

\section{Introduction}
\label{sec:int}

Our goal is to identify functional properties of nodes based on the network structure. Connection between network structure and its functionality is important, many attempts were made to find functional signatures in the network structure, such as \cite{st, ingr}, for a review see \cite{new_s}. As tagging network nodes and edges with functional attributes depends on external information and is not a completely unique procedure, the original problem needs reformulation which is tractable with graph-theoretical tools. 

The function real-world networks perform constrains their structure. Yet, one often has more detailed information about the network structure than about the functions it may perform. We focus on systems, either natural or artificial, which process signals and are comprised of many interconnected elements. From a signal processing point of view, global information about network structure is encoded in the shortest paths, i.e. if signal processing is assumed to be fast, most of network communication is propagated along the shortest paths. Therefore global and local properties of shortest paths are relevant for understanding organisation of the signal processing in the system represented with a suitable network. During  signal transmission, signals are being spread and condensed in the nodes, as well as along network edges. We have previously shown \cite{ejn, prsb} that in case of cerebral cortex, using a simplified version of the convergence degree (CD), it was possible to connect structural and functional features of the network. In complex networks, signal processing characteristics are also determined by the level of network circularity (which in biology and especially neural science is known as reverberation, for obvious reasons). Possibility to go around \textit{chordless} circles necessitates simultaneous quantification of signal condensing, spreading along network edges and edge circularity. Here we generalise edge convergence and divergence \cite{prsb}, and take into account the existence of circles in the network, treating their effects separately from the effect of branching. For that reason we refine the definition of edge convergence and introduce the overlapping set of an edge, both notions are to be defined in a precise manner later in the text. 
Our approach may be viewed as generalisation of in-, out and strongly connected components of a graph to the level of network edges. Notions introduced have an extra gain, they help clarifying the otherwise murky notion of network causality. The functional role of a node in a network is defined by the amount of information it injects to or absorbs from the system, or passes on to other nodes. 
In case of real-world networks we test our findings using external validation, given the existing body of knowledge about each specific network. 
We illustrate the advantage of edge-based approach with the case of strongly connected graphs, where edge-based measures offer deeper understanding of signal processing and transmitting roles of nodes than an analysis which concentrates solely on nodes and their properties. 

Measures we work with are applicable to networks of all sizes, there is no assumption about "sufficient" network size. More precisely, networks we work with can be small, and applicability to large networks is limited only by the computational capacity needed to find all shortest paths in the network. The semantics of our approach is tailored to explain signal flow, though our methodology is applicable to directed networks in general. In cases of information processing, regulatory, transportation or any other network the appropriate semantics of the approach has to be given.  

In Section \ref{sec:notion} we introduce the notions of convergence degree and overlapping set, in Section \ref{sec:nrr} we define the flow representation, in Section \ref{sec:res} we analyse four real-world networks and discuss signal transmission, processing and control properties of the small-world networks. We compute CD-s and (nontrivial) overlap probability distributions for three model networks. In the last section we discuss our results and draw conclusions. 

\section{In-, out and overlapping-sets and the convergence degree}
\label{sec:notion}

Convergence degree was introduced in \cite{prsb} for the analysis of cortical networks and was applied to some random networks \cite{bnnb}. We modify the measure introduced therein, in order to capture the structure of shortest paths in a more detailed way. We will discuss both global and local properties of the shortest paths, relevant notions will be distinguished with self explanatory indices $G$ and $L$ respectively. 

\begin{figure}[!htbp]
\centering
\includegraphics[scale=0.4]{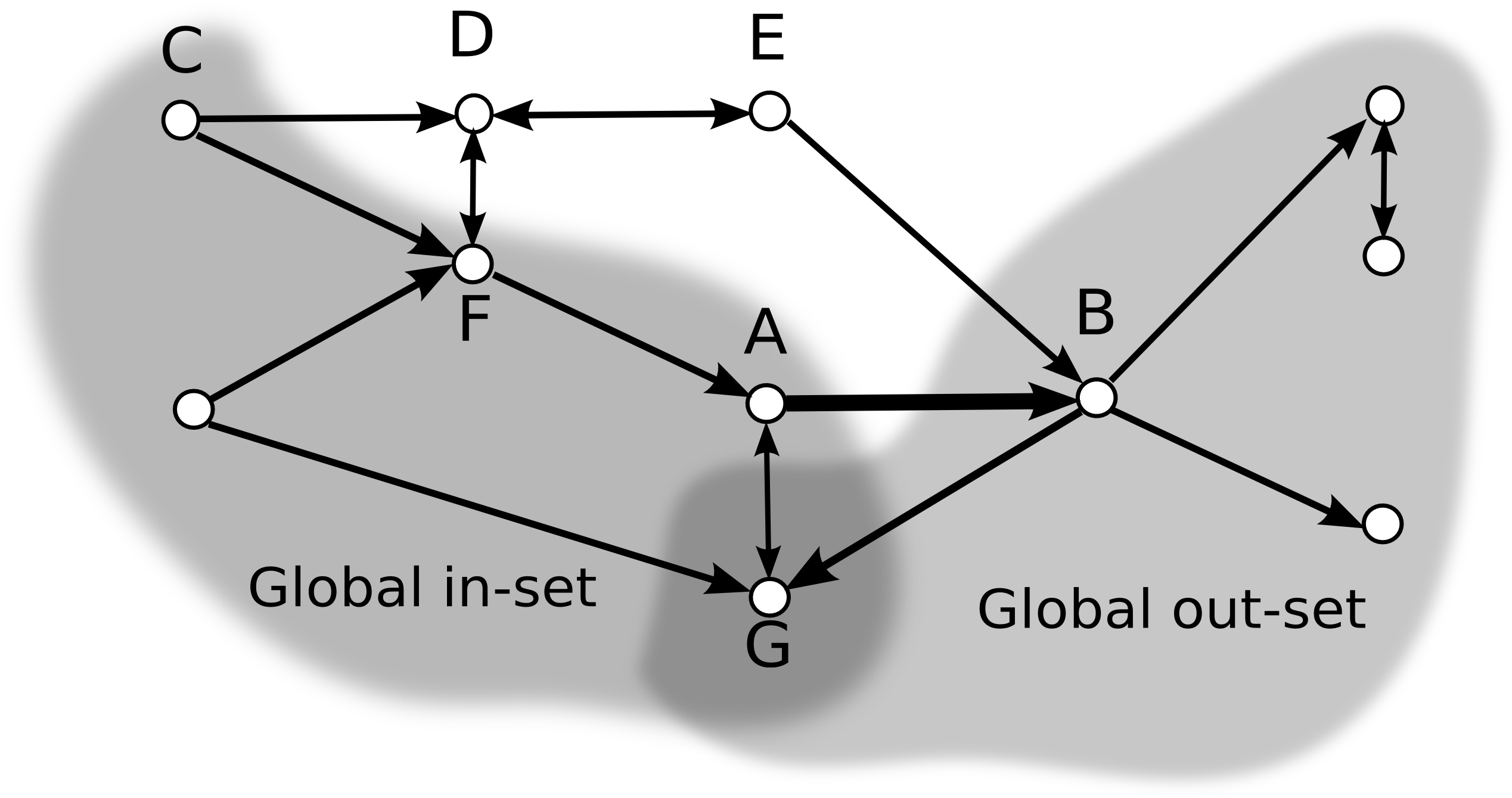}
\caption{In, Out and overlapping sets of the edge $(A,B)$. Global sets are displayed as shaded regions, local sets are comprised of first in-neighbours of node $A$ and first out-neighbours of node $B$ inside the shaded regions, with the exception of node $G$, which is contained in the local and global overlap of $In(A,B)$ and $Out(A,B)$. Note the omition of points $D$ and $E$ from the global input and output sets.}
\label{fig:cd1} 
\end{figure}

Let $SP(G)$ be the set of all the shortest paths in the graph $G$. For any edge $e_{i,j} \in E(G)$ we can choose a subset $SP(G,e_{i,j})$ comprised of all the shortest paths which contain the chosen edge $e_{i,j}$. $SP(G,e_{i,j})$ uniquely determine two further sets: $In_G(i,j)$ the set of all the nodes from which the shortest paths in $SP(G,e_{i,j})$ originate, and $Out_G(i,j)$ the set of all the nodes in which the shortest paths in $SP(G,e_{i,j})$ terminate. By definition we assume that node $i$ is in $In_G(i,j)$ and node $j$ is in $Out_G(i,j)$. We define a third set, $Int(i,j) = In(i,j) \cap Out(i,j)$, the intersection of $In$- and $Out$ sets and call it the overlapping set. We note that $In_G(i,j)$ ($Out_G(i,j)$, respectively $Int_G(i,j)$) is the edge-level equivalent of the in-component (out-component, respectively strongly connected component) of the directed network, introduced in \cite{ioc1} and later refined by \cite{dms}.
Notions relevant for understanding the convergence degree and overlapping set are shown in Figure \ref{fig:cd1}.

\begin{figure}[!htbp]
  \centering
  \includegraphics{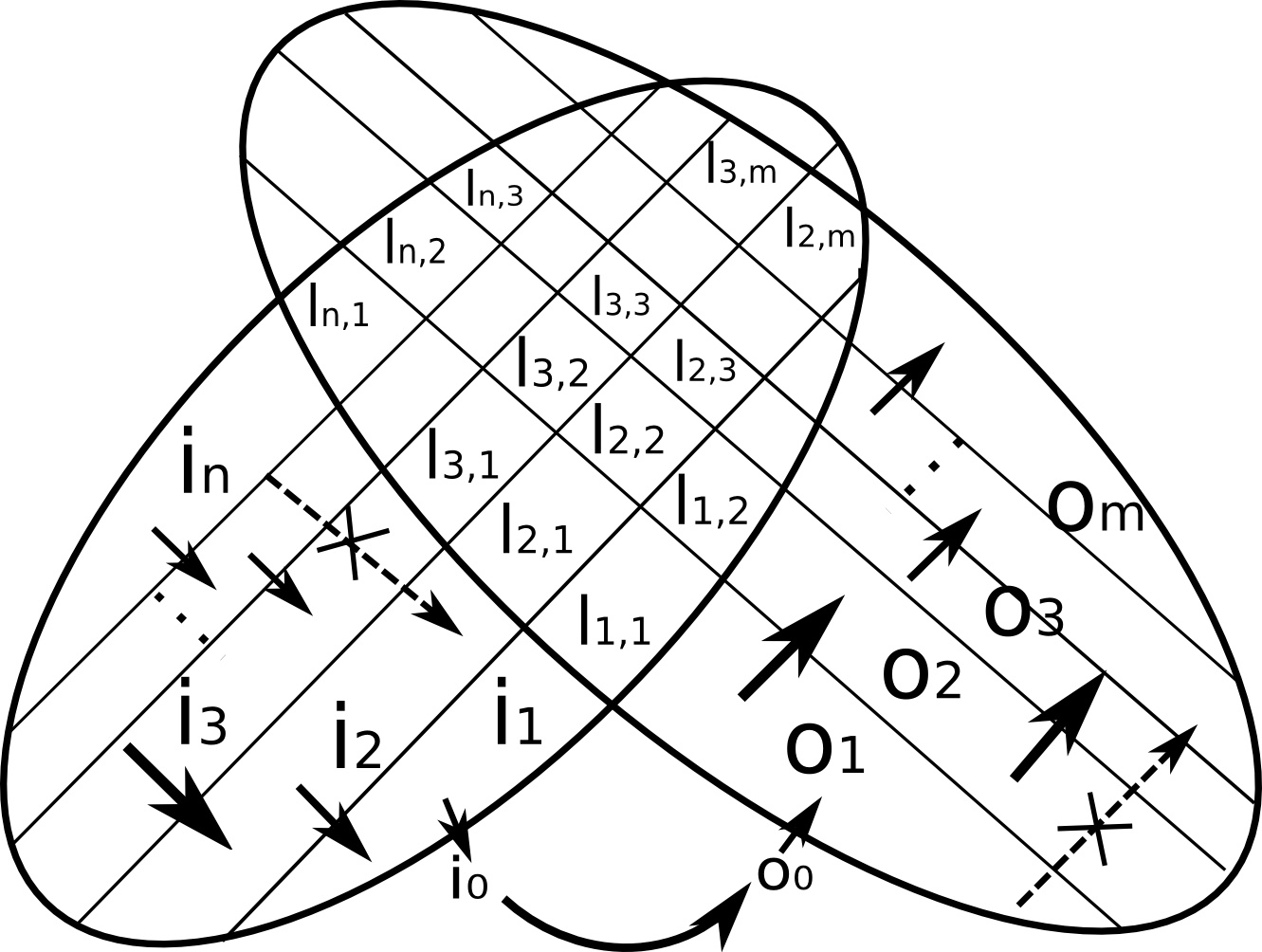}
  \caption{Stratification of global input, output and overlapping sets is shown.Input strata are labelled with indices $i$, output strata are labelled with indices $o$ and overlap strata have double indices $l$. Examples of prohibited edges are shown with dashed lines, necessary edges are shown with full line. Strata $i_0$ and $o_0$ are connected with the edge itself and they do not overlap.} 
  \label{fig:strata}
\end{figure}

From the perspective of the chosen edge, the whole network splits to two, possibly overlapping sets, both of which have rich structure. 
Shortest paths induce natural stratification on the set $In_G(i,j)$, nodes at distance 1, 2 and so on from the node $i$ are uniquely determined. Points at distance $m$ from the tail form the $m$-th stratum of $In_G(i,j)$. 
Each point in the $m$-th stratum is a tail of an edge with a head in the $m-1$-th stratum.  Edges connecting $m$-th stratum with any stratum $n<m-1$ are prohibited. Edges from the $In$ strata to the $Out$ strata are prohibited, since those would alter the shortest paths between the sets. The set $Out_G(i,j)$ is stratified in a similar fashion. Points in the intersection of $In_G(i,j)$ with $Out_G(i,j)$ inherit both stratifications. 
Stratification of $In_G$ and $Out_G$ sets is illustrated in Figure \ref{fig:strata}.

Local versions of these sets are defined as follows: $In_L(i,j)$ is the set of all the first predecessors of the node $i$, while $Out_L(i,j)$ is the set of first successors of the node $j$. When indices $G$ or $L$ are omitted, either is used. If the graph has circles, $In$ and $Out$ sets may overlap, thus it makes sense to introduce strict $SIn$ and $SOut$ sets, which are defined as follows:

\begin{eqnarray}
  \label{eq:SIO}
  SIn(i,j)  & = & In(i,j) \setminus Int(i,j)  \\
  SOut(i,j) & = & Out(i,j) \setminus Int(i,j)
\end{eqnarray}

$In$, $Out$, $SIn$ and $SOut$ are generalisations of the notion of first predecessors and successors of a node, and accordingly, cardinalities of these sets are generalisations of the in- and out-degrees of nodes. We note that global and local versions of the $In$, $Out$ and overlapping sets are two extremes of two set families defined as follows. Let $In(i,j,r_1)$ be the set of points from which paths at distance less or equal to $r_1$ from the point $i$ begin, analogously let $Out(i,j,r_2)$ be the set of points at which paths at distance less or equal to $r_2$ from the point $j$ terminate. The two sets are balls centred at $i$ and $j$ with radii $r_1$ and $r_2$. Instead of balls, one may consider the surfaces of the balls, in which case points at distances $r_1$ and $r_2$ are considered. The global $In$-set is thus $In_G(i,j)=In(i,j,\infty)$, whilst the local $In$-set corresponds to points at surfaces with radii 1, $In_L(i,j)=In(i,j,1)$.

The notion of strict in-, out- and overlapping sets is important for understanding causality relations in network systems. 
Global signal flow through an edge $e_{i,j}$ induces separation of network nodes into four classes: 
\begin{enumerate}
\item $SIn_G(i,j)$, in which are the causes of the flow. 
\item $SOut_G(i,j)$, in which the effects of flow are manifested.
\item The overlap, whose elements represent neither cause nor effect. Relation between elements in the overlap is often described as circular- or network causality. 
\item Points which are not members of $In_G(i,j) \cup Out_G(i,j)$ form the remaining, fourth category which has no causal relationship with the signal flowing through the given edge.
\end{enumerate}
We stress that for a generic graph no such partition is possible based on node properties. E.g. if we tried to define analogous notions based on node properties, all analogue node classes would coincide for the case of strongly connected graphs. The $In$ and $Out$ sets would coincide, and all distinction between different node classes would have been lost.

For each edge we 
define three additional measures, namely the relative size of the strict in-set $(RIn(i,j))$, the relative size of the strict out-set $(ROut(i,j))$, and the relative size of the overlap between in-set and out-set $ROvl(i,j)$, as follows:
\begin{eqnarray}
 RIn(i,j)  & = & \frac{|SIn(i,j)|}{|In(i,j) \cup Out(i,j)|}  \label{eq:UC1} \\
 ROut(i,j) & = & \frac{|SOut(i,j)|}{|In(i,j) \cup Out(i,j)|}  \label{eq:UC2} \\
 ROvl(i,j) & = & \frac{|In(i,j) \cap Out(i,j)|}{|In(i,j) \cup Out(i,j)|}  \label{eq:UC3}
\end{eqnarray}
where $|S|$ denotes the cardinality of the set $S$.

\begin{figure}[!htbp]
  \centering
  \includegraphics{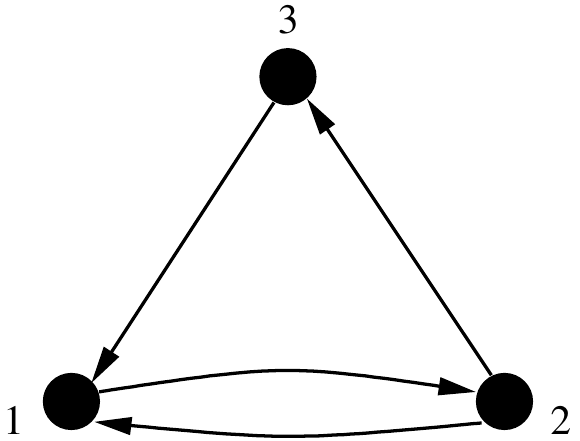}
  \caption{A graph with a chordless circle containing edges with empty and nonempty overlapping sets.}
\label{fig:circ}  
\end{figure}

Note that Equation \ref{eq:UC3} is the Jaccard coefficient \cite{jac2} of the $In(i,j)$ and $Out(i,j)$ sets. 
It is possible to generate networks which have edges with large global overlaps, one simply adds randomly a small number of edges to an initial oriented circle. This example helps understanding the meaning of (possibly large) global overlaps: they are characteristic of edges in chordless circles. More precisely, for and edge to have a nonempty overlapping set it is necessary, but not sufficient, to be on a chordless circle of length at least three. We illustrate this by an example. In the graph shown in Figure \ref{fig:circ}, the only edge with nonempty overlapping set is $e_{1,2}$, with $Int(1,2) = \{3\}$. $e_{1,2}$ is on the chordless circle (3,1,2,3), whilst the edges $e_{3,1}$ and $e_{2,3}$ on the same chordless circle have zero overlapping sets.

Local overlaps are related to the clustering coefficient of the graph, since they define the probability that the vertices in the neighbourhood of a given vertex are connected to each other.

Overlap represents global mutual relationship and a measure of dependence (in terms of chordless circles) between $In$- and $Out$ sets. This dependence is inherent in the network structure. Large Jaccard coefficient of the $In(i,j)$ and $Out(i,j)$ sets is not detectable with edge betweenness, as it may obtain large values for edges with non-overlapping sets.

The edge convergence degree $CD(i,j)$ of the edge $e_{i,j}$ is defined as follows:
\begin{eqnarray}
  \label{eq:cd}
  CD(i,j) = RIn(i,j)-ROut(i,j) =\frac{|SIn(i,j)| - |SOut(i,j)|}{|In(i,j) \cup Out(i,j)|}
\end{eqnarray}
Note that the definition of CD uses the normalised sizes of the strict $In$- and $Out$-sets to make the measure independent of the network size. Furthermore, this formula is related to the complement of the Jaccard coefficient (denoted as $Jacc(\;,\;)$) of the $In$- and $Out$-sets, or equivalently to their normalised set-theoretic difference, thus connecting the CD to information theoretical quantities. The following inequality is obvious:
\begin{eqnarray}
|CD(i,j)| \leq 1 - Jacc(In(i,j),Out(i,j)) = 1 - ROvl(i,j) \label{eq:CD_le_Ovl}
\end{eqnarray}
Directionality of the edge gives meaning to cardinality substraction, as $In$ and $Out$ sets can be distinguished. If the CD value is close to one, the signal flow through the edge is originating from many sources and terminating in very few sinks, while CD values close to -1 indicate flow formed of few sources and many sinks. This property justifies rough division of edges according to their CD properties to convergent (condensing), balanced and divergent (spreading). An oriented circle with at least three nodes has the maximum possible global overlap for each edge, while the absolute value of the global $CD$ is the smallest possible, in accordance with the inequality (\ref{eq:CD_le_Ovl}). We note that CD in an oriented chain monotonously decreases along the chain, whilst the overlap is zero along the chain. This simple example again illustrates how CD and overlap are sensitive to the network topology.

Applicability of the convergence degree is limited by the following facts. Definition of convergence degree makes sense only if not all connections are reciprocal, stated otherwise if there is a definite directionality in the network. If every connection is reciprocal, the network may be considered unoriented. For fully reciprocal networks, the $In$ and $Out$ sets would coincide. Second, convergence degree makes sense for a network which is at least weakly connected.

\section{Flow representation of the network}
\label{sec:nrr}

Since the number of edges exceeds the number of nodes in a typical connected network, and in many cases we are interested in the role of individual nodes, it is desirable to condense the our primarily edge-based measures to a node-centric view. The condensed view should reveal several features of interest: local vs global signal processing properties of network nodes, directionality of the information, i.e. whether we are interested in the properties of the incoming or outgoing edges, the third aspect is the statistics, i.e. total or average property of the edges, and finally we may choose edges according to the sign of their CD. Condensing the information about overlapping sets follows the same lines, with the exception of the sign. 

We proceed by an example and introduce the following six quantities defined for each node $i$. Let $\sigma_{in,L}^{-,av}(i)$ denote the sum of all incoming negative local convergence degrees divided by the node's in-degree, and let $\sigma_{in,L}^{+,av}(i)$ denote the sum of all incoming positive convergence degrees divided by the node's in-degree, i.e. $\sigma_{in,L}^{-,av}(i)$ is the average negative inwards pointing local CD of the node $i$.  

In a similar way we can also define $\sigma_{out,L}^{-,av}(i)$ and $\sigma_{out,L}^{+,av}(i)$ for outgoing convergence degrees. For clarity we give formulae for $\sigma_{in,L}^{-,av}(i)$ and  $\sigma_{out,L}^{-,av}(i)$. $d_{in}(i)$ and $d_{out}(i)$ denote in-degree and out-degree of the node $i$, $\theta$ is the unit step function continuous from the left. $\Gamma_{in}(i)$ denotes the first in-neighbours of the node $i$, the analogous notation $\Gamma_{out}(i)$ is selfexplanatory. 

\begin{eqnarray}
 \sigma_{in,L}^{-,av}(i) & = & \frac{1}{d_{in}(i)} \sum_{j \in \Gamma_{in}(i) } \theta(-CD_L(j,i)) CD_L(j,i) \\
 \sigma_{out,L}^{-,av}(i) & = & \frac{1}{d_{out}(i)} \sum_{j \in \Gamma_{out}(i)} \theta(-CD_L(i,j)) CD_L(i,j)
\end{eqnarray}

We also define $\sigma_{in,L}^{ovl,av} (i)$, the sum of all incoming local overlaps and $\sigma_{out,L}^{ovl,av} (i)$, the sum of all outgoing local overlaps each being normalised with the corresponding node degree. 
\begin{eqnarray}
 \sigma_{in,L}^{ovl,av}(i) & = & \frac{1}{d_{in}(i)} \sum_{ j \in \Gamma_{in}(i) }  ROvl_L(j,i) \\
 \sigma_{out,L}^{ovl,av}(i) & = & \frac{1}{d_{out}(i)} \sum_{j \in \Gamma_{out}(i) }  ROvl_L(i,j)
\end{eqnarray}
Factors before the sums serve normalisation purposes, each $\sigma$ should have a value within the $[-1,1]$ interval. 
These quantities are average local CD-s and relative overlaps corresponding to each node. One is also interested in the total of the in- and out pointing edges of a given CD sign, and define the corresponding version of the node-reduced convergence degree. For normalisation purposes the sums in $\sigma^{tot}$'s are divided by $n-1$, the maximal possible number of the outgoing (incoming) connections a node can have, where $n$ denotes the number nodes in the network.

Thus, using the quantities $ \sigma_{\{in, out\}, \{G, L\}}^{\{+,-\},\{tot ,av\}}$  and $\sigma_{\{in, out\}, \{G, L\}}^{ovl, \{tot ,av\}}$ one can construct four different CD flow representations of a network, namely $CD_{G}^{tot}$, $CD_{G}^{av}$, $CD_{L}^{tot}$ and $CD_{L}^{av}$.

The incoming node-reduced CD values are understood as coordinates of the $x$ axis, while the outgoing CD values are interpreted as the coordinates of the $y$ axis. In order to display overlaps together with the convergence degrees in a single figure, overlaps are treated as the coordinates of the $z$ axis, the incoming overlaps being positive and the outgoing understood negative. Each point is represented in each octant of the flow representation. The points in the $xy$ plane are not independent, given the values in the diagonal quadrants, the other two quadrants can be reconstructed with  reflections.

Representation of graph nodes in the $xy$ plane is related to the CD flow through the nodes in the following way. The CD flow $\phi$ through the node $i$ is defined as follows: 
  \begin{eqnarray}
    \phi(i) = \sum_{j=1}^{d_{out}(i)} CD(i,j) - \sum_{j=1}^{d_{in}(i)} CD(j,i) 
  \end{eqnarray}
The first sum is equal to $\rho_{out}(i) \left(\sigma_{out}^{+} + \sigma_{out}^{-}\right)$, where $\rho(i)$ is the appropriate weight, whilst the second sum equals $\rho_{in}(i) \left( \sigma_{in}^{+} + \sigma_{in}^{-}\right)$. The flow can be rewritten as
\begin{eqnarray}
  \phi(i) = \rho_{out}(i) \sigma_{out}^{+}(i) - \rho_{in}(i) \sigma_{in}^{-}(i) + \rho_{out}(i) \sigma_{out}^{-}(i) - \rho_{in}(i) \sigma_{in}^{+}(i) \label{eq:flow_2}
\end{eqnarray}
If the first difference on the right hand side of Equation (\ref{eq:flow_2}) is large (small), i.e. the representative point is close to the diagonal $y=-x$ and is far from the origin in the top left (bottom right) quadrant, and the second difference is small (large), i.e. the representative point is close to the diagonal $y=-x$ and is far from the origin in the bottom right (top left) quadrant, the node $i$ is \textit{source} (\textit{sink}) of the CD flow. 
Analogously, the CD flow can be written as:
\begin{eqnarray}
  \phi(i) = \rho_{out}(i) \sigma_{out}^{-}(i) - \rho_{in}(i) \sigma_{in}^{-}(i) + \rho_{out}(i) \sigma_{out}^{+}(i) - \rho_{in}(i) \sigma_{in}^{+}(i) \label{eq:flow_3}
\end{eqnarray} 
where the two differences determine the router characteristics of the node $i$. In this sense flow representation is a means to independently study different components of the CD flow. Different circles may have common nodes, thus the overlap flow defines whether different circles passing through the given node have more common parts after of before the given node, i.e. whether a node is a source or sink of circularity. Precise meaning of large and small depends on the criteria used to classify the representative points of the node-reduced representation. 

\begin{figure}[!htbp]
\centering
\includegraphics[scale=0.7]{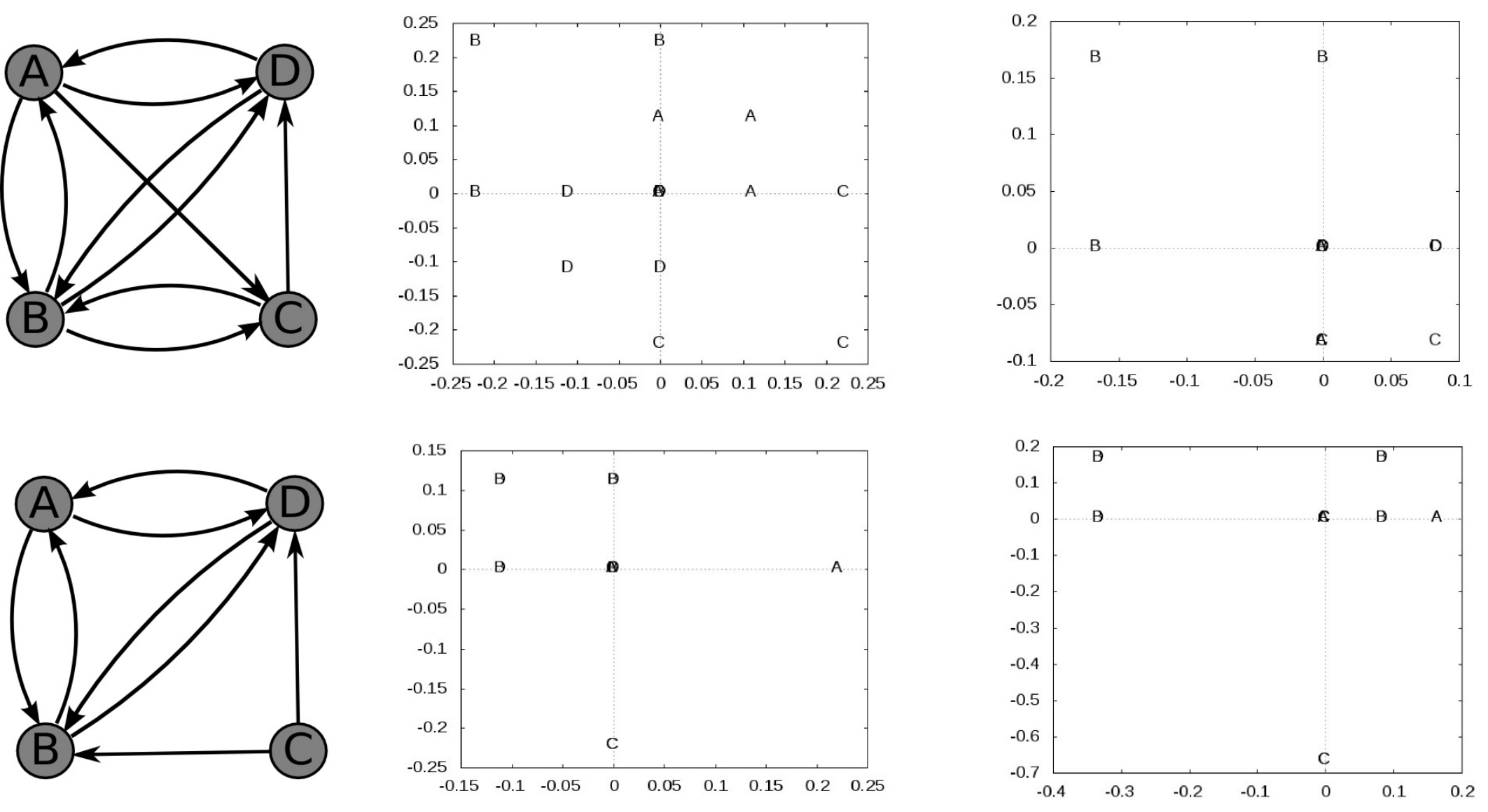}
\caption{The lower graph differs in two edges from the top graph. The middle column represents graph nodes with $\sigma_{G}^{tot}$, the right column represents graph nodes with $\sigma_{L}^{tot}$. Every overlapping set is empty for the lower graph, because all chordless circles are of length two. Some points have the same coordinates in the flow representation. E.g., point D is is global, but not local CD flow sink.}
\label{fig:lepke} 
\end{figure}

Nodes can be classified based on the CD (relative overlap) flow, besides distinction based on the sign, the scale is continuous, there is no a-priori grouping of nodes. Further classification can be made based on the structure of the CD (relative overlap) flow, i.e. based on properties of different terms defining the CD (relative overlap) flow. Components of the flow representation for two toy graphs are shown in Figure \ref{fig:lepke}. We observe that same nodes may be global, but not local CD flow sinks or sources.
Each octant represents different aspect of convergence-divergence relations in the network. These quantities bring us to the actual interpretation of edge convergence and divergence as a characterisation of signal flow on the nodes of a network. To make statements about the signal flow derived from the CD flow, we have to make an inversion of properties, as nodes which behave as a sink of convergence, actually inject information to the network, thus they are sources of signal. Respectively, CD sources are sinks of signal. Assuming this interpretation we can extract useful information from the flow representation regarding the signal processing roles of nodes in the network. Nodes which have incoming edges with cardinalities of the $In$sets ($Out$sets) being larger than cardinalities of the $Out$sets ($In$sets), and outgoing edges with cardinalities of the $Out$sets ($In$sets) being larger than cardinalities of the $In$sets ($Out$sets) are, from the signal processing perspective, identified as sources of signals. The combination of divergent input (negative incoming CD sum) and convergent output (positive outgoing CD sum) is, considering the signal flow, equivalent to absorption of signals in the network. This is represented in the top left quadrant of the $xy$ plane. On the opposite, the combination of convergent input and divergent output corresponds to the source characteristics of the nodes (bottom right quadrant of the $xy$ plane). The top right and bottom left quadrants can be interpreted as a display of \textit{signed} relay characteristics of the nodes. Nodes which have incoming edges with cardinalities of the $Out$sets ($In$sets) being larger than cardinalities of the $In$sets ($Out$ets), and outgoing edges with cardinalities of the $Out$sets ($In$sets)being larger than cardinalities of the $In$sets ($Out$sets), are called negative (positive) router nodes. At the same time routing characteristics can be read from the top right and bottom left quadrants. Routers \textit{redistribute} incoming CD of a given sign to outgoing CD of the \textit{same} sign. Additional information is obtained from the $z$ coordinate, which gives the average overlap of incoming and respectively, outgoing edges. This quantity identifies the degree of a node's participation in signal circulation in the network, a property typically associated with control circuits.

Graphical presentation of a network is not unique, e.g. isomorphic graphs may look totally different, the Petersen graph being a typical example. Community structure is not unique, grouping of points, thus presenting a network can be achieved in a multitude of ways. Yet, the flow representation of a network is \textit{unique}, though due to possible symmetries it may have a significant amount of redundancy. This 3D plot of the network is unique in the sense that there is no arbitrariness in the position of the points in the three dimensional space. The flow representation can be considered as a network fingerprint since isomorphic graphs are mapped to the same plot, and differences between flow representations can be attributed to structural and functional properties of the network. If all edges are reciprocal or the graph is undirected, the flow representation of the network shrinks to a single point. The same argument applies to all graphs in which some nodes can not be distinguished due to symmetries. More precisely, nodes in the orbit of an element generated by the automorphism group of the graph are represented with the same point on the flow representation, as all the value of $\sigma$-s are constants on the orbits generated by the automorphism group of the graph. 

Usefulness and application of the flow representation will be illustrated in the analysis of the real-world networks in Section \ref{sec:real_net}.

\section{Results}
\label{sec:res}

We calculate CD-s for three model networks and analyse CD-s of 
four real-world networks.

\subsection{Signal flow characteristics of real-world networks}
\label{sec:real_net}

In this section we analyse functional clusters in real-world networks and the statistical properties of their interconnection. We analysed two biological and two artificial networks: macaque visuo-tactile cortex \cite{ejn, prsb}, signal-transduction network of a CA1 neuron \cite{ma'a}, the call graph of the Linux kernel version 2.6.12-rc2 \cite{lkern}, and for comparison purposes the street network of Rome \cite{roma}. Nodes and edges are defined as follows: in the macaque cortex nodes are cortical areas and edges are cortical fibres, in the signal-transduction network nodes are reactants and edges are chemical reactions, in the call graph nodes are functions and edges are function calls, in the street network the nodes are intersections between roads and edges correspond to roads or road segments. The first three networks perform computational tasks, Linux kernel manages the possibly scarce computational resources, signal-transduction network can be considered as the operating system of a cell, while cortex is an ubiquitous example of a system which simultaneously performs many computationally complex tasks. The street network is an oriented transportation network, which has a rich structure, as its elements have traffic regulating roles. 

The call graph of the Linux kernel was constructed in the following way. We created the call graph of the kernel source which included the smallest number of components necessary to ensure functionality. The call graph was constructed using the CodeViz software \cite{codeviz}, but it was not identical to the actual network of the functions calling each other, because the software detects only calls that are coded in the source and not the calls only realized during runtime. The resulting call graph had more than $10^4$ vertices. As we wanted to perform clustering and statistical tests, the original data was prohibitively large, therefore we applied a community clustering algorithm \cite{lat} to create vertex groups. We generated a new graph in which the vertices represented the communities of the original call graph and have added edges between vertices representing communities whenever the original nodes in the communities were connected by any number of edges. Definition of the call graph nodes and their connections is analogous to the nodes and connections of the cortical network, as millions of neurons form a cortical area, and two areas are considered to be connected if a relatively small number of neurons in one area is connected to a small number of neurons in another area. The call graph of the Linux kernel will be discussed in Section \ref{sec:aggreg_netw}. 

\begin{figure}
\centering
\includegraphics[scale=0.85]{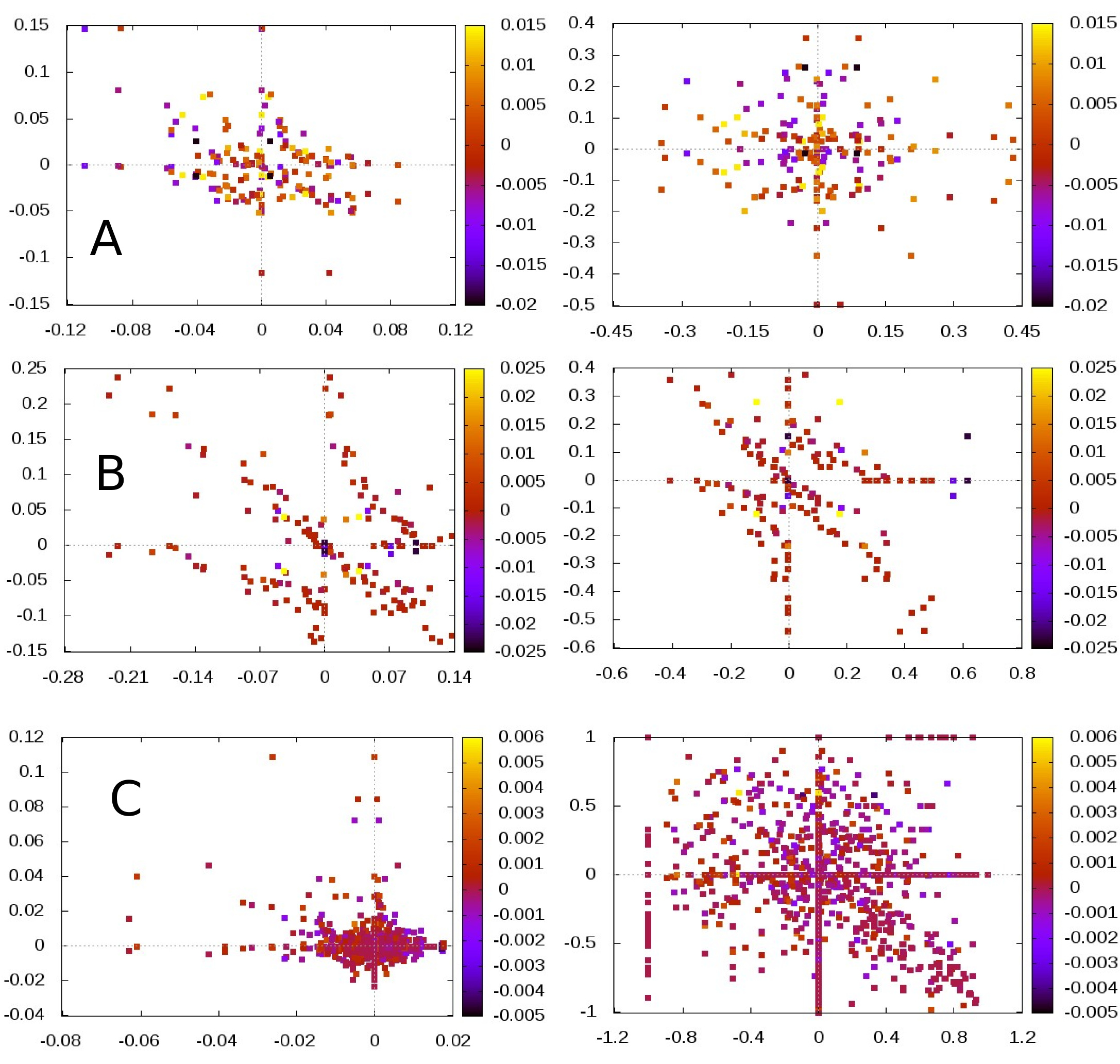}
\caption{Components of the total $CD_G$ flow are shown in the left column, components of the average $CD_L$ are shown in the right column.  
Displayed are: Erd\H{o}s-R\'enyi graph (row A), macaque visuo-tactile cortex (row B) and signal-transduction (row C). Relative overlap flow is indicated by colour intensity.}
\label{fig:3lepke} 
\end{figure} 

The flow representations of two real-world networks are shown in Figure \ref{fig:3lepke} and for comparison, in part A, the Erd\H{o}s-R\'enyi network. We can identify the most important nodes and some general features of the networks as follows. Part B refers to the macaque visuo-tactile cortex. It is characterised by the alignment of the nodes along a straight line along the main diagonal, and hyperbolic-like pattern in the first and third quadrants, showing reverse ordering in the opposite quadrants, and absence of routers, which refers to a hierarchical organisation. In part C one can see the signal-transduction network of a hippocampal neuron. In the signal-transduction network of the hippocampal neurons, the molecules with the most negative CD flow are involved, among other functions, in the regulation of key participants of the signal transduction cascade such as the cAMP second messengers. Molecules with large positive CD flow play function in cell survival and differentiation, as well as apoptosis. Router-like proteins are involved in diverse functions, notably the regulation of synaptic transmission in addition to those mentioned above. However, it should be noted that partly because of the paucity of our knowledge about many of the components of this network, as well as because of redundancy, i.e. overlapping functionality, we could give here only a very superficial classification. All edges of the signal transduction network fall in one of the three classes: excitatory, inhibitory and neutral, \cite{ma'a}. CD and overlap data were unrelated to the inhibitory, excitatory or neutral nature of network edges. 
Empirical distributions of CD-s and overlaps were alike for each edge class, see Figure \ref{fig:stn_edges} in the Appendix. 

\subsubsection{Comparison of local and global structural organisation}

We have analysed the flow representations in order to identify different features of signal processing. Network nodes are points represented in a 6D space of the flow representation, and in order to identify different signal processing, transmitting and controlling groups of nodes we performed clustering using Gaussian mixture and Bayesian information criterion implemented in R \cite{R}. We wish to stress that the clustering we performed is not a form of community detection, but grouping of nodes with respect to their functional signal processing properties. Community detection can identify dense substructures, but it provides no information about the nature of signal processing, transmission or control. In each network we determined local and global, total and average signal processing clusters, have determined their properties, and have analysed the nature of CD-s and relative overlaps within and between clusters.

Clustering of nodes with respect to their functional properties resulted in contingency tables, with clusters being labels of the contingency table, and entries in the contingency able being numbers of edges within and between respective clusters. To estimate the randomness of the contingency tables we performed Monte Carlo implementation of the two sided Fisher's exact test. Number of replicates used in the Monte Carlo test was $10^4$ in each case. The exact Fisher's test characterises the result of the clustering procedure, it quantifies how much the distribution of edges within and between clusters differ. We summarise the results in Table \ref{tab:eft}. For comparison purposes benchmark graphs were generated using algorithms described in \cite{fort-lanch}. 

\begin{table}[!htbp] 
\caption{Number of functional clusters ($n$) and the corresponding $p$-values calculated using Fisher's exact test of the contingency tables. Q denotes the modularity of the community structure. Two numbers in a single cell denote the first two moments derived from sample size of 100 graph instances. Networks are denoted as follows: VTc -  macaque visuo-tactile cortex, stn - signal-transduction network of the hippocampal CA3 neuron, kernel - call-graph of the Linux kernel, Rome - Rome street network, ER - Erd\H{o}s-R\'enyi graphs and bench - benchmark graphs. Numbers were rounded to minimise the table size. Definitions of aggregated networks are given in Section \ref{sec:aggreg_netw}.}
\begin{center}
 \begin{tabular}{|l|c|c|c|c|c|c|} \hline
   network & $n_{comm}$ &Q & $n_{G,tot}$ & $p_{G,tot}$    & $n_{L, av}$ & $p_{L,av}$ \\ \hline
    VTc     & 4    & 0.332 & 6    & 0.48     &  9     &   $10^{-4}$ \\ \hline
    stn     & 58   & 0.530 & 3    & 0.75     & 19     & $10^{-4}$  \\ \hline
    Rome    & 39   & 0.907 & 18   & $10^{-4}$ & 19     & $10^{-4}$ \\ \hline
    ER      & 3.68 & 0.114 & 3.94 & 0.59     &  5.39  & 0.66 \\ 
            & 1.55 & 0.020 & 2.34 & 0.30     &  3.34  & 0.29  \\ \hline
    benchm  & 3.19 & 0.449 & 3.83 & 0.19     &  5.21  & 0.10  \\ 
            & 0.50 & 0.042 & 2.07 & 0.24     &  3.36  & 0.20 \\ \hline \hline \hline
   kernel aggr.  & 18   & 0.426 & 12   & 0.41     & 19     & 0.40 \\ \hline
   stn aggr.  &	9         & 0.34    &	18      &  0.38      &           7 & 0.05      \\ \hline
   Rome aggr. &	6         & 0.46    &	 8      &  0.24      &           5 & 0.86      \\ \hline
  \end{tabular}
\label{tab:eft}
\end{center}
\end{table}
Based on Table \ref{tab:eft}, classification of nodes according to their functional properties does not match the network community structure. Classifying nodes according to their local and global functional properties differ substantially, further details are given in Table \ref{tab:eft_2}. 
The $p$-values of the global and local groupings differ in the same way for all the networks analysed, though the difference is much smaller or absent for call graph of the Linux kernel. Distribution of edges between different node clusters measured by total $CD_G$ flow in the signal transduction network was highly irregular, whilst very regular according to other flow measures. We note that the sizes of overlapping sets, and also the circularities were largest in the signal transduction network, which was a consequence of edge sparseness. Measured by all the $p$-values, the street network had very regular structure, and was distinctively different from all other networks. In the case of Erd\H{o}s-R\'enyi graphs there was practically no difference in randomness between local and global functional clusters, as presence of any community or structure in these networks was a matter of pure chance. Erd\H{o}s-R\'enyi and benchmark networks were parametrised to match the macaque visuo-tactile network. The number of communities was comparable, but the number of functional clusters and the way in which edges connected functional clusters was different. The Erd\H{o}s-R\'enyi and benchmark graphs were both structureless, but in different way.  As one would expect, Erd\H{o}s-R\'enyi graphs had much more randomness in the connectional pattern between functional clusters than the benchmark graphs. In the macaque visuo-tactile network the connection according to the total $CD_G$ was highly irregular, and resembled the Erd\H{o}s-R\'enyi graph, according to other measures the connectional pattern between functional clusters was regular, and differed from the either Erd\H{o}s-R\'enyi or benchmark graphs. Summarising, the $CD_{G}^{tot}$ flow representation is well suited to distinguish properties of signal and information processing networks and captures the characteristical features of signal transmission, processing and control. 

\subsubsection{Analysis of aggregated networks}\label{sec:aggreg_netw}

The amount of data comprised in large networks necessitates community level understanding of signal flow. 
Communities themselves perform signal transmission, processing and control tasks, therefore determination of community level functional properties based on structural information poses a relevant problem. Number of communities in the street network and the hippocampal signal transduction network was large enough to define a nontrivial aggregated network which was subject of analysis. Each community in the original network was represented by a node in the aggregated network. Nodes of aggregated networks had additional structure, namely members of communities they represented, therefore allowing analysis relating CD and overlap flow with nodal structure. 

\begin{figure}
  \centering
 \includegraphics[scale=0.32]{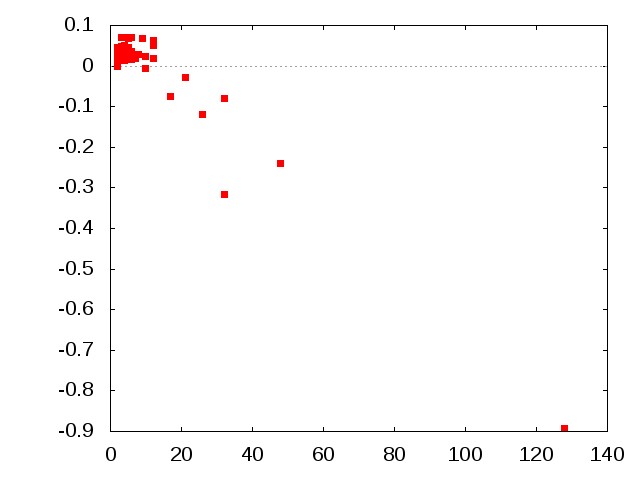}
 \includegraphics[scale=0.32]{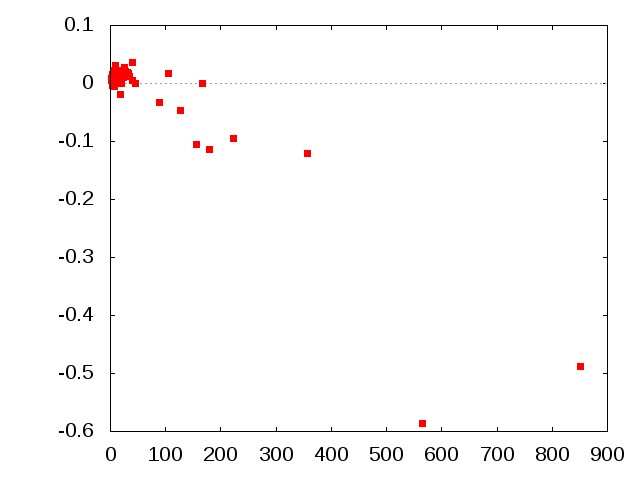}
  \caption{Relation between $CD_G$ flow (vertical axis) of the node in the aggregated network and the cluster size (horizontal axis) in the original network. Results for the signal transduction network is shown in the left panel, results for the Linux call graph are given in the right panel.}
 \label{fig:CD_cl_s}
\end{figure}

\begin{figure}[!htbp]
  \centering
  \includegraphics[scale=0.5]{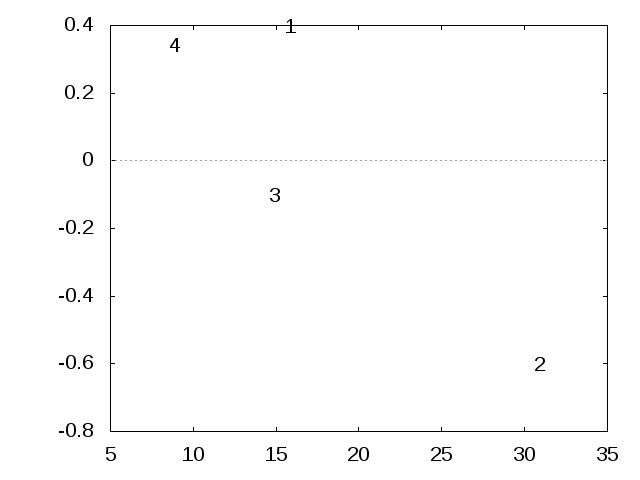}
  \caption{Relation between $CD_G$ flow (vertical axis) of the node in the aggregated network and the cluster size (horizontal axis) in the macaque cortex. The four communities are: 1 - visual related, 2 - higher cognitive functions, temporal, parietal prefrontal and hippocampal formation, 3 - sensory-motor related, 4 - auditory related.}
\label{fig:mc_aggr}
\end{figure}

The $CD_G$ flow of the aggregated networks showed a regular pattern, nodes with positive $CD_G$ flow were numerous and corresponded to small sized clusters in the original network, whilst nodes with negative $CD_G$ flow were few and corresponded to large clusters in the original network, see Figure \ref{fig:CD_cl_s}. 
With some precaution (because of small network size and many unknown edges) analogous analysis of the whole macaque cortical network \cite{mcn} can be performed. The aggregated network had four nodes, see Figure \ref{fig:mc_aggr}. Node with the largest negative $CD_G$ flow corresponded to areas related to higher cognitive functions, the visual and auditory communities were smaller and had positive CD flows. Sensory-motor community had small negative CD flow, and was of intermediate size. 

Similar analysis of the circularity flow revealed that nodes which corresponded to largest clusters in the original network had circularities close to zero. Because in- and out circularities of nodes corresponding to large clusters were nonzero, these nodes were well nested within chordless circles in the network. This nesting enables efficient performance of control-related tasks.  CD flows of the original networks were mainly positive in the nodes corresponding to small, positive CD flow clusters. At the same time, only in nodes representing large clusters which had negative CD flow were numerous nodes with negative CD flows.
Given the different nature of networks analysed, we conclude that organising principles in large-scale networks manifest dependence of functional roles on sizes of the network communities. 

\begin{figure}
 \centering 
 \includegraphics[scale=0.85]{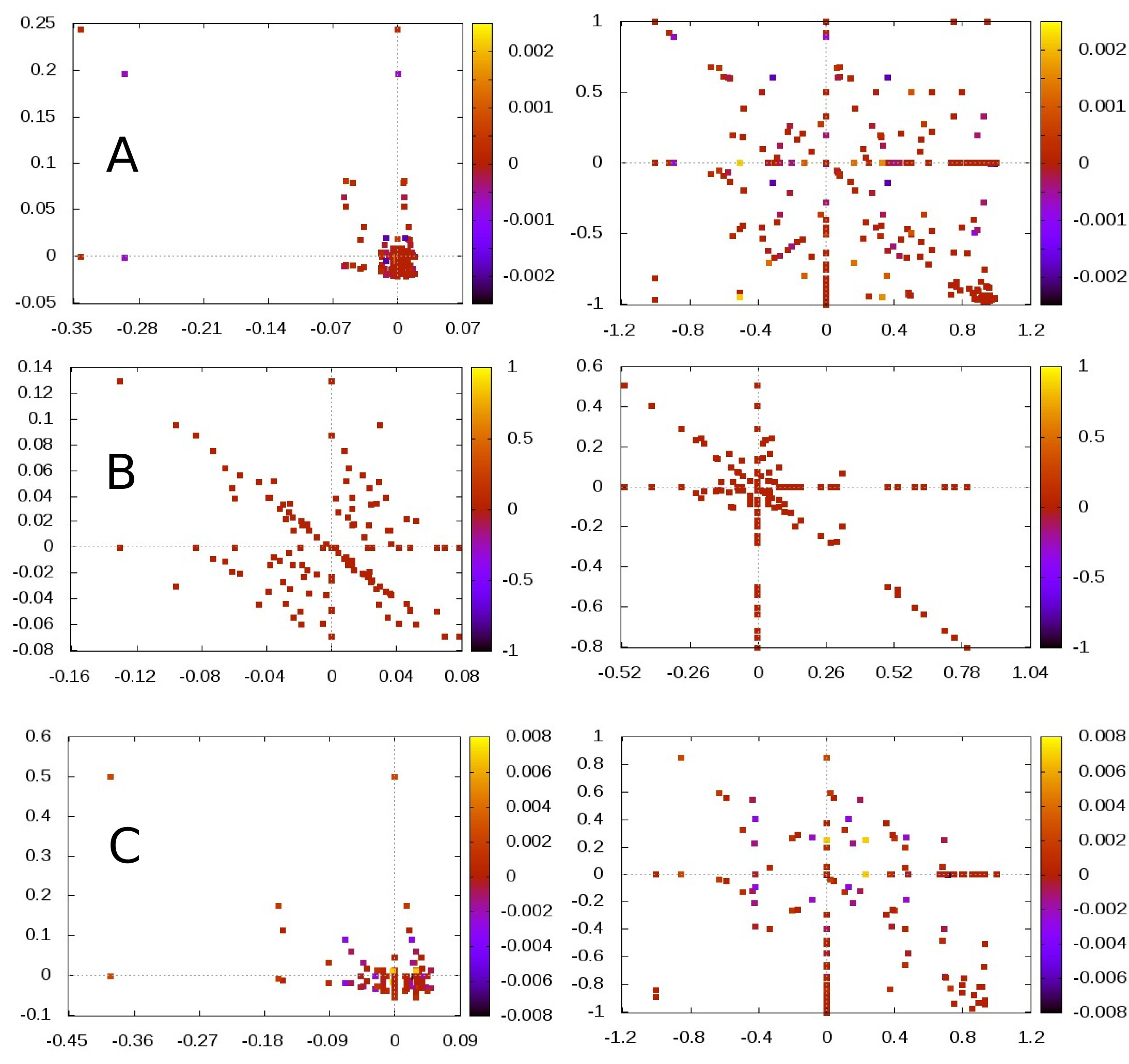}
  \caption{Flow representation of the aggregated networks. Components of the total $CD_G$ flow are shown in the left column, components of the average $CD_L$ flow are shown in the right column. 
Displayed are: Linux call graph (row A), street network (row B) and hippocampal signal transduction network (row C). Relative overlap flow is indicated by colour intensity.}
  \label{fig:aggreg}
\end{figure}

In case of the Linux call graph the most outlying nodes in the CD flow representation are the memory initialisation and buffer operators as CD flow sources, some of the CD flow sink nodes are connected to file system operations and the task scheduler. 
Flow properties of the aggregated street- and hippocampal signal transduction networks differ from the original networks, and resemble the properties of the macaque visuo-tactile cortex, as shown by aggregation of points along the $y=-x$ line in the diagonal quadrants, and grouping of points in the other two quadrants, see Figure \ref{fig:aggreg}. This is a signature of different organisation principles of signal transmission, processing and control properties at the community level, the net CD on the incoming side of a node is roughly redistributed on the outgoing side with a change of sign.  

Statistical results of the analysis of functional properties were summarised in the lower part of Table \ref{tab:eft}. Randomness of connections between functional clusters in the aggregated street network strikingly differs from the original street network. Functional properties of the aggregated signal transduction network are similar to the functional properties of the cortical network, measured by the $p$-values. A possible explanation is that communities, i.e. functional cellular compartments of the signal transduction network have much better defined functional roles than single units, thus from the functional point of view, the role of nodes in the aggregated network is comparable to the cortex, when cortex is represented as a network of cortical areas. 

\subsubsection{Signal flow in small-world-like networks}

Small-world property is often mentioned in relation to cortical (and other) networks. As CD- and overlap-related properties describe important features of signal transmission, processing and control, we studied whether signal flow properties can be obtained by the small-world generating algorithms. Macaque visuo-tactile cortex is strongly connected, even more, it contains numerous Hamilton circles. We constructed and analysed random graphs which matched prescribed properties of the cortical network. 

The Watts-Strogatz graphs were generated as follows: we started from a directed circle. Then we added edges sampling the source and target vertices from uniform distribution until we reached the desired edge count. If the reciprocity was preset, after each new edge with the probability defined by the reciprocity, we added an edge from the target to the source vertex as well. When the preferential algorithm was applied, the distribution of the source and target vertices were sampled as defined by the out- and in-degrees of the vertices respectively. This meant that a higher degree induced a proportionally higher probability for the vertex to be chosen as source or target. For statistical comparison we generated 100 graph instances of each network. Some numbers were rounded, in order to optimise the table size. 

We used Kolmogorov-Smirnov test to check whether CD-s and relative overlaps of the cortical and generated graphs originated from the same (statistically indistinguishable) probability density function. For each instance of generated graph the answer was negative. Statistical results are shown in Table \ref{tab:gen_stat}.

\begin{table}[!htbp]\label{tab:CD_stat}
\caption{ER denotes Erd\H{o}s-R\'enyi graph, sw denotes small-world, swp denotes small-world with preference, VTc denotes macaque visuo-tactile cortex. All networks were of the same size, $|V(G)|=45$, $|E(G)|=463$, and the proportion of the reciprocal edges was 0.8. Two numbers in a cell are the values of the first two empirical central moments, with the exception of Kolmogorov-Smirnov test results, where they denote $D$ and $p$ values respectively.}
\begin{center}
  \begin{tabular}{|l|c|c|c|c|c|c|c|}\hline 
   netw.     & clust.             & diam.             & average                  & CD                    & KS-test      & ROvl              &   KS-test    \\ 
             & coeff.             &                   & short. path          &                        & $D$, $p$      &                  &   $D$, $p$   \\ \hline
   ER        & 0.550              & 3.1               & 1.88                 & $2 \cdot 10^{-3}$      & 0.28         & 0.11             &   0.81       \\ 
             & $2\cdot 10^{-3}$    & $3\cdot 10^{-2}$   & $3\cdot 10^{-3}$     &  0.26                   & 0          & 0.075            & 0            \\ \hline
   sw        & 0.600              & 3.06              & 1.89                 & $\;$ $2\cdot 10^{-3}$   & 0.089         & $5\cdot 10^{-3}$  & 0.047        \\
             &  $1\cdot 10^{-3}$   & $2\cdot 10^{-2}$   & $3\cdot 10^{-3}$     &  0.54                   &  0.11         & 0.03             &  0.66        \\ \hline 
   swp       & 0.623              & 4.32              & 1.93                 & $\;$ $1.6\cdot 10^{-2}$ & 0.096         &  $5\cdot 10^{-3}$ &    0.046     \\ 
             & $3\cdot 10^{-3}$    & $8\cdot 10^{-2}$  & $7\cdot 10^{-3}$       &  0.64                  & 0.10          & 0.03             &  0.299       \\ \hline 
   VTc       & 0.517              &       5           &    2.15               & $\;$ $2\cdot 10^{-2}$   &               & $8\cdot 10^{-3}$ &              \\ 
             &                    &                   &                       & 0.57                   &               & 0.45             &              \\ \hline
\end{tabular}
\end{center}
\label{tab:gen_stat}
\end{table}

We conclude that description of cortical networks as small-world networks can be only a qualitative statement, as the small-world model fails to capture features relevant from the signal processing, transmission and control perspective.  

\subsection{Model networks}
\label{sec:mod_net}

It is possible to calculate the Cd-s and overlaps or their probability density functions for some networks. 

\subsubsection{Arborescences}

The purpose of calculating CD for arborescences is the comparison with networks grown with preferential attachment mechanism, see Section \ref{sec:pa}. We calculate global convergence degree of a complete directed tree  -- sometimes called arborescence. We assume that the root is at level 0, the number of levels is $n$, the branching ratio is constant and equals $d$ and that all the edges are directed outwards from the root. For clarity, with the exception of the root, all in-degrees are equal to 1, and with the exception of the leaves, all out-degrees are equal to $d$. If all assumptions are true, between any pair of nodes there is either no shortest path or there is only one. At level $k$ $(0\leq k \leq n)$ the cardinality of any $In$ set is $k$, while at level $k+1$ the size of any $Out$ set is the sum of a geometric progression: $\frac{d^{n-k}-1}{d-1}$. Thus with some abuse of notation $CD_G$ of any edge connecting nodes at levels $k$ and $k+1$ equals:

\begin{eqnarray}
  \label{eq:CD_fa}
  CD_G(k,k+1) = 1 - \frac{2}{1+\frac{k(d-1)}{d^{n-k}-1}}
\end{eqnarray}

We observe that edges originating from the root have negative convergence degrees, but as the level index increases soon there are two possibly distinct levels $k_1$ and $k_2$, such that for $k \leq k_1$ $CD_G$ is negative, whilst for $k \geq k_2$ $CD_G$ is positive. $k_1$ and $k_2$ may coincide, or $k_2 = k_1 + 1$. $k_1$, and $k_2$ are determined by the solution of the equation $d^{n-k} + k(d-1) =1$. Thus almost all edges have positive convergence degrees. One would na\"{\i}vely expect that all the edges in such a tree are divergent, yet most of them are not. There is a level at which the number of the nodes in the $In$ and $Out$ sets results in the exchanged order of their (relative) sizes. The overall convergence in the whole network gives:

\begin{eqnarray}
 N(n,d) & = &\sum_{k=0}^{n-1} d^k CD_G(k,k+1) > 0 \label{eq:net_CD_fa} 
\end{eqnarray}

Calculation of the local convergence degree is trivial: 

\begin{eqnarray}
  CD_L(k,k+1) = \frac{1-d}{1+d}, \quad CD_L(n-1,n) = 1
\end{eqnarray}

Contrary to the global CD there is only a trivial change in sign of the local CD. 

\subsubsection{Preferential attachment networks}\label{sec:pa}

Based on \cite{bnk} we calculated the CD probability density function for the network grown with preferential attachment mechanism. This network has the structure of a random tree, therefore all overlapping sets are empty. 

In growing networks it is natural to orient all the edges towards the root. For stratified networks, based on \cite{bnk} one can derive local and global CD probability density function of nodes at distance $n$ from the root, i.e. nodes at $n$-th level of the network. According to \cite{bnk} the degree distribution at the level $n$ is given as 
\begin{eqnarray}
  f^{(n)}(k)  = (1+y)\frac{\Gamma(2+y) \Gamma(k)}{\Gamma(2+k+y)} \label{eq:fnk}
\end{eqnarray}
where $y$ is the depth measured in units of average depth: 
\begin{equation}
  \label{eq:y}
  y=\frac{n-1}{\langle n-1 \rangle}
\end{equation}

Let $x$ denote the $CD_L$ of an edge connecting levels $n+1$ and $n$. 
\begin{equation}
  \label{eq:x}
  x= \frac{k_{n+1} - 1}{k_{n+1} + 1}
\end{equation}
where $k_{n+1}$ denotes the in-degree of the node at level $n+1$. Probability density of the local CD is calculated by changing the variable in Equation (\ref{eq:fnk}) according to Equation (\ref{eq:x}). The  probability density of local CD having value $x$ for an edge between levels $n+1$ and $n$ is:
\begin{eqnarray}\label{eq:pxn}
  P_L(x,n)= \frac{2}{(1-x)^2} \, f^{(n+1)} \left(  \frac{1+x}{1-x} \right)
\end{eqnarray}
Let $g^{(n)}(s)$ denote the probability of finding a tree rooted in the $n$-th layer of size $s$. $g^{(n)}(s)$ can be written as follows, \cite{bnk}:
\begin{equation}
  \label{eq:gns}
  g^{(n)}(s) = \frac{1+y}{2+y} \, \frac{\Gamma\left( 2+\frac{y}{2} \right)}{\Gamma\left(\frac{1}{2}\right)} \, \frac{\Gamma \left( s - \frac{1}{2} \right)}{\Gamma \left( s+1+\frac{y}{2} \right)}
\end{equation}
Let $x$ denote the random value of the global CD for an edge connecting levels $n+1$ and $n$. 
\begin{eqnarray}
  x=\frac{s_{n+1}-n}{s_{n+1}+n} \label{eq:p_fa}
\end{eqnarray}
where $s_{n+1}$ denotes the fact that it is described with $g^{(n+1)}$. After changing the variable in (\ref{eq:gns}), according to Equation (\ref{eq:p_fa}), the probability density of the global CD for an edge connecting layers $n+1$ and $n$ is:
\begin{equation}
  \label{eq:CDpdf}
  P_G(x,n)=\frac{2 n }{(1-x)^2} \, g^{(n+1)}\left( n \, \frac{1+x}{1-x} \right)
\end{equation}
From the last term in the numerator of the Equation (\ref{eq:gns}) one concludes that the domain of $P_G$ is the open interval $\left( \frac{1-2n}{1+2n}, 1\right)$, which is the probabilistic equivalent of the global CD sign change observable in arborescences.

\subsubsection{Erd\H{o}s-R\'enyi graphs}

Calculation of the CD and relative overlap probability density is based on the fact that all relevant probabilities are related to binomial distribution or a distribution derivable from a binomial one. Closed formulae for the local CD and overlap probability density function can be given, though they are lengthy, see Equations (\ref{eq:pL_CD}, \ref{eq:pL_ROvl}). In the global case, the exact PDF are given by a recursive formula of considerable depths. 

Calculation of CD-s for Erd\H{o}s-R\'enyi graphs is straightforward, though lengthy. We note that the Erd\H{o}s-R\'enyi graphs \cite{er} we work with are \textit{directed}. Furthermore for clarity we note that loop edges and multiple edges are prohibited. First we calculate the probability density function of $CD_L$, if number of nodes is $n$ and the probability of having an edge between any two nodes is $p$. Let $i$ denote the in-degree of the tail of the edge, let $o$ denote the out-degree of the head of the same edge, and let $l$ denote the number of nodes in the intersection of the first in-neighbours and out-neighbours of the tail and the head of the given edge. There are two essential terms in formulae below. The first is the one defining how large is the set of nodes we can choose our actual set from, the upper term in the binomial coefficients. The second one is the one defining which edges are prohibited to have the actual set size, the exponents in the $(1-p)$ terms. The exponent of the $p$ terms and the lower terms of the binomial coefficients are simply the sizes of the node sets we choose. The probability of an edge tail having $i$ predecessors is given with binomial density function: 
\begin{eqnarray}
  p(i) = 
\left( 
\begin{array}{c}
  n-1 \\
  i
\end{array}
\right)
p^i (1-p)^{n-1-i} \label{eq:p_i}
\end{eqnarray}
The probability of an edge head having $o$ successors is given with Equation (\ref{eq:p_i}), with $i$ replaced with $o$.

The probability of having an intersection of the predecessors of the tail and the successors of the head of size $l$, given the size of the input and output sets, can be calculated as follows. First, if we assume that $i=o=l$, the probability $p_{l}^{*}$ of having an overlap of size $l$ is given as follows: 
\begin{eqnarray}
  p^{*}(l) = 
\left( 
\begin{array}{c}
  n-1 \\
   l
\end{array}
\right)
p^{2l} (1-p)^{2(n-1-l)} \label{eq:p*}
\end{eqnarray}

We can take into account the non-overlapping parts of the input and output sets as follows, where the conditional probability of $l$ given $o$ (ranging from $l$ to $n$) and $i$ (ranging from $l$ to $n-o$) is:
\begin{eqnarray}
  p(l|i,o) & = & p^{*}(l) 
\left(
\begin{array}{c}
  n-1-l \\
   o -l
\end{array}
\right) 
p^{o-l} (1-p)^{n-1-o} 
\left(
\begin{array}{c}
  n-1-o -l \\
    i-l
\end{array}
\right)
p^{i-l} (1-p)^{n-1-o-i} \label{eq:p(l|i,o)}
\end{eqnarray}

Let $p(i,o,l)$ denote the joint probability density function of the variables $i, o$ and $l$, it can be given as: 
\begin{eqnarray}
  p(i,o,l) = p(l|i,o) p(i,o) = p(l|i,o) p(i) p(o) \label{eq:p_er_joint}
\end{eqnarray}
We note that in Equation (\ref{eq:p_er_joint}) $i$, $o$ and $l$ can be chosen independently, with $l$ ranging from $0$ to $\min(i,o)$. 
The value of $CD_L$ is given as $(i-o)(i+o-l)^{-1}$. We perform the change of random variables 
\begin{eqnarray}
\psi(i,o,l)=(x,y,z), \quad x=\frac{i-o}{i+o-l}, \quad y=o, \quad z=l. \label{eq:cv_cd}
\end{eqnarray}
Changing the variables in the probability density function given with Equation (\ref{eq:p_er_joint}) and calculating the marginal probability results in probability density function for $CD_L$:
\begin{eqnarray}
  p(x) = \sum_{y, z =1}^{n-1} p\left( \frac{x(z-y)-y}{x-1} , y, z \right) \frac{\left|z-y\right|}{(x+1)^2} \label{eq:pL_CD}
\end{eqnarray}
Similarly, to obtain $p_O$, the probability density function of the relative size of the overlapping set, one proceeds with the following change of variables:
\begin{eqnarray}
\psi(i,o,l)=(x,y,z), \quad x=i, \quad y=o, \quad z=\frac{l}{i+o-l} \label{eq:cv_ovl}
\end{eqnarray}
and ends up with the following the probability density function:
\begin{eqnarray}
  p_O(z) = \sum_{x, y =1}^{n-1} p\left( x, y, \frac{(x+y)z}{1+z} \right) \frac{x+y}{(z+1)^2} \label{eq:pL_ROvl}
\end{eqnarray}

Calculation of probability density function for $CD_G$ is recursive. Nodes in the input set are organised into strata according to their distance from the edge head, the cardinalities of the strata being $i_k$, $k$ ranging from 0 to $n-1$, thus the cardinality of the input set is given as:
\begin{eqnarray}
  i=\sum_{k=0}^{n-1} i_k  \label{eq:i}
\end{eqnarray}
When calculating $CD_G$ edges are allowed to the stratum $i_{s-1}$ and all other shortcut edges from stratum $i_s$ to lower strata are prohibited, including head and tail of the edge whose $CD_G$ we are interested in. Loop edges are also prohibited. Strata in the output set are analogously denoted as $o_s$, meaning the $s$-th stratum in the output set. We bistratify the overlapping set, so its cardinality can be calculated  in the following way:
\begin{eqnarray}
  l=\sum_{i \leq j} l_{i,j} \label{eq:l}
\end{eqnarray}
where $l_{ij}$ denotes the overlap of the $i$-th stratum of the input set with the $j$-th stratum of the output set. We note that with probability 1 the cardinality of zeroth stratum in the input and output set is 1. Also, from the definition of zeroth strata it follows $l_{0,0}=0$ with probability 1. \\
To shorten the subsequent formulae we use the following notation:
\begin{equation}
I_k=\sum_{r<k}i_r, \quad O_k=\sum_{r<k}o_r, \quad L_{a,b}=\sum_{r<a} \: \sum_{r \leq m < b} l_{r,m}
\end{equation}
Probability of having $i_s$ nodes in the $s$-th stratum is:
\begin{eqnarray}
  p(i_s|i_{s-1}, \dots, i_0) & = & \sum_{a=i_s}^{n-1-I_{s}}
\left(
\begin{array}{c}
  n-1 - I_{s} \\
  a
\end{array}
\right) 
a \sum_{j=1}^{i_{s-1}} p^j (1-p)^{n-1+I_{s-1}} \label{eq:p_i_th_strat}
\end{eqnarray}
We note the restriction on values $i_s$ may have: $0 \leq i_s \leq n - I_s$. The conditional probability in Equation (\ref{eq:p_i_th_strat}) was calculated according to the following lines. 

The dummy variable $a$ indicates the number of nodes at in-distance $s$ from the tail of the chosen edge. The limit of the first summation is the same term as the upper expression in the binomial coefficient, represents the number of available nodes to choose the $m$-th stratum from. The summation and multiplication by $a$ before $p^j$ accounts the fact that every node in the $s$-th stratum of the In-set can be attached to any number of nodes in the $s-1$-th stratum. The $I_{s-1}$ term in the exponent of $p-1$ represents the prohibition of edges from the $s$-th stratum to the lower strata except for the one right below it. The complementary term for $p^j$ would be $(1-p)^{n-1-j}$, but the $-j$ in the exponent is compensated by the prohibition of edges to the tail of the given edge from all points of the $s$-th stratum. All subsequent formulae are derived using similar reasoning. \\
According to the definition of the conditional probability, we have

\begin{eqnarray}
  p(i_s, \dots , i_0) = p(i_s | i_{s-1}, \dots i_0) \dots p(i_1|i_0) p(i_0) \label{eq:p(i)}
\end{eqnarray}

Probabilities of $o_k$-s are calculated analogously, with $i$ replaced by $o$, and $a$ replaced by $b$ denoting the number of nodes at outdistance $s$ from the head of the chosen edge. \\
Calculation of the conditional probability of having an overlap of size $l$ is recursive. As nodes in the overlapping set share properties of the input and output sets, exponent of the $(1-p)$ term has to prohibit all shortcuts which are prohibited from both sets. \\
The analogue of Equation (\ref{eq:p*}) is:
\begin{eqnarray}
p^*(l_{s_1, s_2} | i_{s_1},i_{s_1 -1}, \dots , i_0 ; o_{s_2}, o_{s_2 -1}\dots , o_0; l_{s_1-1, s_2}, \dots , l_{0, 0}) & = & \nonumber \\
&&\hspace*{-10cm}\sum_{a=l_{s_1,s_2}}^{n-1-L_{s_1,s_2}}\sum_{b=l_{s_1,s_2}}^{n-1-L_{s_1,s_2}}
\left( 
\begin{array}{c}
  n-1 - L_{s_1,s_2} \\
  a+b-l_{s_1,s_2}          
\end{array}
\right) 
a b \sum_{j_1=1}^{i_{s_1-1}} \sum_{j_2=1}^{o_{s_2-1}} p^{j_1 j_2} (1-p)^{n-1+I_{s_1-1}+O_{s_2-1}} 
\label{eq:p*lio} 
\end{eqnarray}

Possible values of $l_{s_1, s_2}$ in Equation (\ref{eq:p*lio}) are restricted as follows: $0 \leq l_{s_1, s_2} \leq \min(i_{s_1}, o_{s_2})$. The conditional probability of having excess over the overlap in the output set is given as:
\begin{eqnarray}
p^{\%} (l_{s_1, s_2} | i_{s_1}, \dots , i_0; o_{s_2}, \dots , o_0; l_{s_1-1, s_2}, \dots , l_{0, 0}) & = & \nonumber \\
&&\hspace*{-6cm} 
\sum_{a=l_{s_1,s_2}}^{n-1-L_{s_1,s_2}-O_{s_2}}
\left(
\begin{array}{c}
  n-1-L_{s_1,s_2}-O_{s_2} \\
  a
\end{array}
\right) 
a \sum_{j=1}^{o_{s_2-1}} p^j (1-p)^{n-1+O_{s_2-1}}
\end{eqnarray}

Analogously, the conditional probability of the input set being larger than the overlap is:
\begin{eqnarray}
  p^{\#} (l_{s_1, s_2} | i_{s_1}, \dots , i_0; o_{s_2}, \dots , o_0; l_{s_1-1, s_2}, \dots , l_{0, 0}) & = & \nonumber \\
&&\hspace*{-8cm}\sum_{b=l_{s_1,s_2}}^{n-1-L_{s_1,s_2}-O_{s_2}-I_{s_1}}
\left(
\begin{array}{c}
  n-1-L_{s_1,s_2}-O_{s_2}-I_{s_1} \\
  b
\end{array}
\right) 
b \sum_{j=1}^{i_{s_1-1}} p^j (1-p)^{n-1+I_{s_1-1}}
\end{eqnarray}

The conditional probability of $l_{s_1 s_2}$ (global analogue of Equation (\ref{eq:p(l|i,o)})) is given as: 
\begin{eqnarray}
p (l_{s_1, s_2} | i_{s_1}, \dots , i_0; o_{s_2}, \dots , o_0; l_{s_1-1, s_2}, \dots , l_{0, 0}) & = & \nonumber \\
&&\hspace*{-5cm}p^{*} (l_{s_1, s_2} | i_{s_1}, \dots , i_0; o_{s_2}, \dots , o_0; l_{s_1-1, s_2}, \dots , l_{0, 0}) \nonumber \\
&&\hspace*{-5cm} p^{\%} (l_{s_1, s_2} | i_{s_1}, \dots , i_0; o_{s_2}, \dots , o_0; l_{s_1-1, s_2}, \dots , l_{0, 0}) \nonumber \\
&&\hspace*{-5cm} p^{\#} (l_{s_1, s_2} | i_{s_1}, \dots , i_0; o_{s_2}, \dots , o_0; l_{s_1-1, s_2}, \dots , l_{0, 0})
\end{eqnarray}

Thus, analogously to the Equation (\ref{eq:p_er_joint}), using Equation (\ref{eq:p(i)}) and its analogue for the output set, the joint probability of $i_{s_1}$, $o_{s_2}$ and $l_{s_1 s_2}$ is:
\begin{eqnarray}
  p_J(i_{n-1}, \dots , i_{0}, o_{n-1}, \dots , o_0, l_{n-1, n-1}, \dots , l_{i_0, o_0}) & = & \nonumber \\
&&\hspace*{-4cm}\prod_{k_1, k_2 = 0}^{n-1}
p (l_{k_1, k_2} | i_{k_1}, \dots , i_0; o_{k_2}, \dots , o_0; l_{k_1-1, k_2-1}, \dots , l_{0, 0}) \label{eq:p_joint}
\end{eqnarray}

Based on Equations (\ref{eq:p_joint}, \ref{eq:i}, \ref{eq:l}) one derives the marginal probability function $p_M(i,o,l)$ (which is the global analogue of Equation (\ref{eq:p_er_joint})), with $0 \leq l \leq \min(i,o)$:
\begin{eqnarray}
  p_M(i,o,l) & = & \sum_{s_1=0, \dots , s_{n-1}=0 }^{n-1, \dots , n-1} \quad
                   \sum_{t_1=0, \dots , t_{n-1}=0 }^{n-1, \dots , n-1} \quad
                   \sum_{u_{11}=0, \dots , u_{n-1 \: n-1}=0 }^{ s_1+t_1, \dots , s_{n-1}+t_{n-1}} \nonumber \\ 
&&\hspace*{-2.8cm}p_J \left( x_{s_0}, x_{s_1} - x_{s_0}, \dots, i - x_{s_{n-1}}, y_{t_0}, y_{t_1} -y_{t_0}, \dots, o - y_{t_{m-1}}, u_{0,0}, \dots, l- u_{n-1, n-1} \right)
\end{eqnarray}
then proceeds with the change of variables given in Equations (\ref{eq:cv_cd}), and calculates the marginal probability of $x$ resulting in $CD_G$ probability density of the same form as the one given in Equation (\ref{eq:pL_CD}). $p_O$, the probability density function of the relative size of the overlapping set is calculated using the change of variables given in Equations (\ref{eq:cv_ovl}), in $p_M (i,o,l)$. Finally, one obtains the probability density function of the same form as the one given in Equation (\ref{eq:pL_ROvl}).

\section{Discussion}
\label{sec:disc}

\begin{figure}[!htbp]
\centering
\includegraphics[scale=0.33]{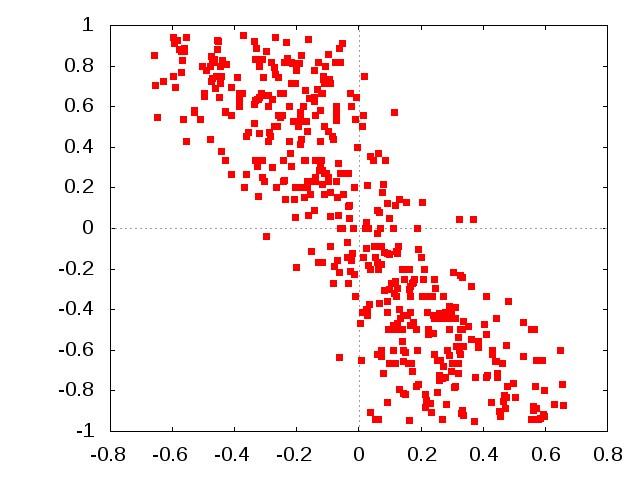}
\includegraphics[scale=0.33]{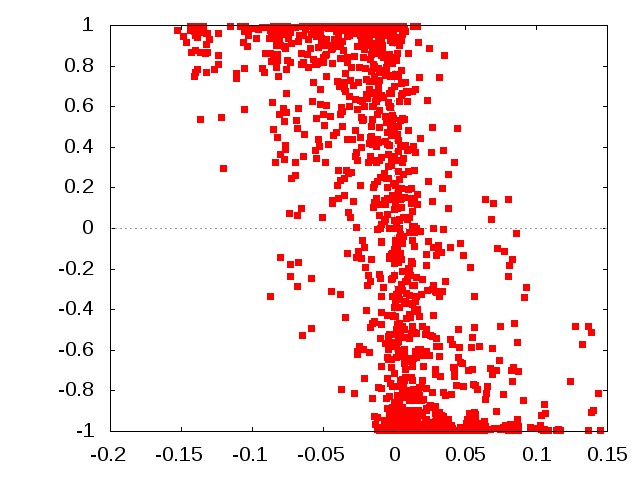}
\caption{Relation between the CD flow through the nodes at the ends of am edge and CD of the same edge, points displayed have $(\phi_j - \phi_i, CD(i,j))$ coordinates. Data is shown for the cortical- (left panel) and hippocampal signal transduction network (right panel).}
\label{fig:struct}
  \end{figure}

Octants in the flow representation allow study of hierarchical organisation in the network, as flow sink nodes are assumed to be at lower hierarchical positions than the flow source nodes, \cite{prsb, luo}. Flow sink nodes are connected with flow source nodes via edges with negative CD values, usually identified as feed-forward connections, while flow source nodes are connected to flow sink nodes via edges with positive CD, usually identified as feed-back connections, see Section \ref{sec:real_net}. 
More precisely, based on graph structure it is possible to define a partial order relation on the set of nodes $V(G)$. Node $i$ precedes node $j$ according to the CD (ROvl) flow relation $\geq_{CD \; (ROvl)}$ if and only if $\phi_i > \phi_j$, where $\phi$ denotes the CD (ROvl) flow. In terms of hierarchical flow (HF) \cite{luo}, $\geq_{HF} \equiv \leq_{CD}$. 
The consistency of classification edges as feed-forward or feed-back based on structural information is formulated as a 
relation between the CD flow through a node and the CD of edges attached to a node, and is shown in Figure \ref{fig:struct}, where the values of CD plotted against the difference of CD flows of the nodes at the two ends of and edge. The feed-forward or feed-back nature of edges could be verified using background information on the networks under study. 
As our analysis of the real-world networks have shown, notions of convergence degree and overlapping sets may serve as initial steps in the task of relating a network's structure and functional properties it may have.   

From the functional perspective, properties of the convergence degree and overlap can be understood as follows. 
Signals propagating through a given edge originate from the $In$-set, and are received in the $Out$-set. At the same time, signals are not simply transmitted or processed, as many real-world networks perform control tasks. 
Traditionally, in case of biological networks edges were classified as feed-forward and feed-backward and parts of control architecture were understood in such terms. We argue that such approach can be complemented with the introduction of simplest control loops. 
The basic building blocks of control systems are comprised of chordless circles. Overlapping set and circularity grasp some properties of the control systems inherent in the network structure. The methodology introduced relies on the notion of shortest paths. Many real-world networks have large number of non shortest paths, for example to ensure fault tolerance. It is possible that not all the signals are transmitted along the shortest paths. The effect of non shortest paths can be grasped without introducing dynamics. Our methodology can be extended in principle to answer how the functionality of network elements is altered. One may work with paths exceeding the length of shortest paths by one, and from the set of all such paths for each edge  define the $In$ and $Out$ multisets, and proceed as we did. The procedure can be iterated if necessary.

Analytical description of CD was given for two tree-like networks. Absence of circles in trees results in CD properties which are different from other networks. Knowledge of consequences the presence of circles on CD may have are important for understanding the role of circulation, thus control in signal processing and transmission in real-world networks. Various properties of special graph classes are often compared to Erd\H{o}s-R\'enyi graphs in statistical tests. 
It was possible to determine the CD and overlap probabilities for the Erd\H{o}s-R\'enyi graphs, because they have a special property, statistical homogeneity, yet real-world networks are nonhomogenous. Whether further graph properties allow at least approximate calculation of CD probabilities remains to be seen. Asymptotic expressions of relevant probability distributions describing Erd\H{o}s-R\'enyi graphs are highly desirable. 

Our analysis of CD and overlap flows can be interpreted in terms of information flow and circulation. Identification of routers, sinks, sources and circulating nodes in the real-world networks was in accordance with the known functional roles of the nodes, for related previous work see \cite{prsb}. Control and other loops were already investigated, \cite{ma'a} and classified as positive or negative depending on the nature of edges (excitatory or inhibitory) they contained. Our methodology allows identification of an edge being feed-forward or feed-back in terms of CD flow and offer another definition of positive or negative feed-back loops. In the neuronal signal transduction network feed-forward and feed-back nature of an edge was independent from an edge being excitatory, inhibitory or neutral. Previous work concentrated on control-related motives which were subnetworks of relatively small size. In contrast, our methodology in its extreme can focus on the whole network. Analysis of aggregated networks revealed connection between functional properties of communities and their size. A possible explanation is that communities performing integrative tasks are highly specialised, and are comprised of relatively small number of elements. 
Communities performing allocatory and control related tasks perform broader class of more general tasks and are therefore comprised of larger number of elements. Allocation and control is centralised in the sense that the number of communities performing such general tasks is relatively small. 

Functional roles and their interrelations are neither exact, nor sharp, they are rather tendencies observable after a suitable form of information reduction. Our treatment of the flow representation resembles the phenomenological approach of \cite{amaral}, as nodes are represented in appropriate space, but the space in which we represented the nodes and the way in which nodes were grouped differed substantially. Our analysis had three further gains: clarification of the network causality, demonstration of importance of chordless circles and a fresh look to the small-world characterisation of networks. Small-world property is important and is defined with a generating algorithm which has a clear intuitive meaning. Yet contrasting small-world networks (generated using standard generating algorithms or their combination) with the cerebral cortex revealed that they had different CD and overlap statistics. 

The cortical network has no pronounced routers, which fact may be related to the evolutionary process that optimised signal processing in the brain for speed. Evolution may also explain the lack of the nodes which only pass signals. Cortex preserved only the minimum number of nodes necessary for performing all the computational steps, i.e. every signal transmission is inseparable from signal processing. We demonstrated similar organisation in other aggregated networks. 

Our study of the Linux kernel call graph was far from complete, further analysis and inclusion of runtime calls will refine our interpretation of particular nodes at a finer scale. Deeper analysis of the neural signal-transduction network is likely to shed further insight into the low level signal transmission and processing of the cortex.

It was shown that signal processing, transmitting and controlling properties of a given network depend on the definition of a node. By aggregating a community into a single node and applying the same methodology, one can explore signal transmission and processing at the community level. Aggregated networks had different properties from the original networks, thus coarsening the network unit resolution revealed very different community-level information processing, transmitting and control properties. Further analysis of the real-world networks will be given elsewhere.

In signal and information processing networks global functional organisation was much more random than the local one. This means that global and local organisation principles differ, and stochasticity may play a role on the large scale, while local connectivity is functionally more constrained. 

The reason for global functional randomness can be understood as follows. Different processing streams have nodes with similar functional properties, though these properties are exercised over different domains, as it was shown for the cerebral cortex \cite{prsb}. There is no general rule which would require connection between different integrator nodes in different domains, say. When there is such a connection it is likely to be an important one. 

We have also shown a real-world example of a transportation network, which had markedly different properties from the signal processing networks. The finding is not based on comparison of structural, but rather functional properties. This was an example of how the nature of the network constrains its functional organisation. 

Our goal was to understand the influence of structure on the functional properties of networks. A dynamic complex network model would consist of two main objects, the temporal processes and a space where these processes take place. The tools and methods in this paper only address the description of the network as a static object, contributing to the definition of the discrete nonhomogenious space of a dynamic network model. Further research is needed to understand dynamic features of information convergence and divergence, including the analysis of temporal processes taking place on networks.

\section{Appendix}

\subsection{Statistical analysis of functional organisation}

For sake of completeness in Table \ref{tab:eft_2} we complement Table \ref{tab:eft} with further results of statistical analysis. 

\begin{table}[!htbp]
  \caption{Networks coincide with those of Table \ref{tab:eft}. Shown are omitted entries, two numbers in a cell are the first two empirical moments.}
\begin{center}
  \begin{tabular}{|c|c|c|c|c|c||c|c|c|} \hline 
   network     & VTc     & stn      &  Rome     & ER    & benchm & kernel aggr & stn aggr & Rome aggr  \\ \hline
   $n_{G,av}$   & 9       &      8   &  19       & 3.9   & 4.3    & 12          &  7       & 8 \\ 
               &         &          &           & 2.47  & 2.58   &             &          &  \\ \hline
   $p_{G,av}$   &0.03     & $10^{-4}$ &  $10^{-4}$ & 0.62  & 0.23   & 0.08         & 0.18     & 0.53 \\ 
               &        &           &           & 0.32  & 0.26   &              &          & \\ \hline
   $n_{L, tot}$ & 9       & 15       & 14         &  4.64 & 5.14   & 10           & 5        & 23 \\ 
              &         &          &            &  2.99  & 3.02  &              &          &   \\ \hline
   $p_{L, tot}$ & $10^{-4}$& $10^{-4}$ & $10^{-4}$  &  0.61 & 0.10   & 0.14         &0.04      & 0.93 \\ 
              &         &           &           &  0.28 & 0.21   &              &          &       \\ \hline
  \end{tabular}
\end{center}
  \label{tab:eft_2}
\end{table}

\subsection{CD and overlaps of the neural signal transduction network}

\begin{figure}[!htbp]
  \centering
  \includegraphics[scale=0.6]{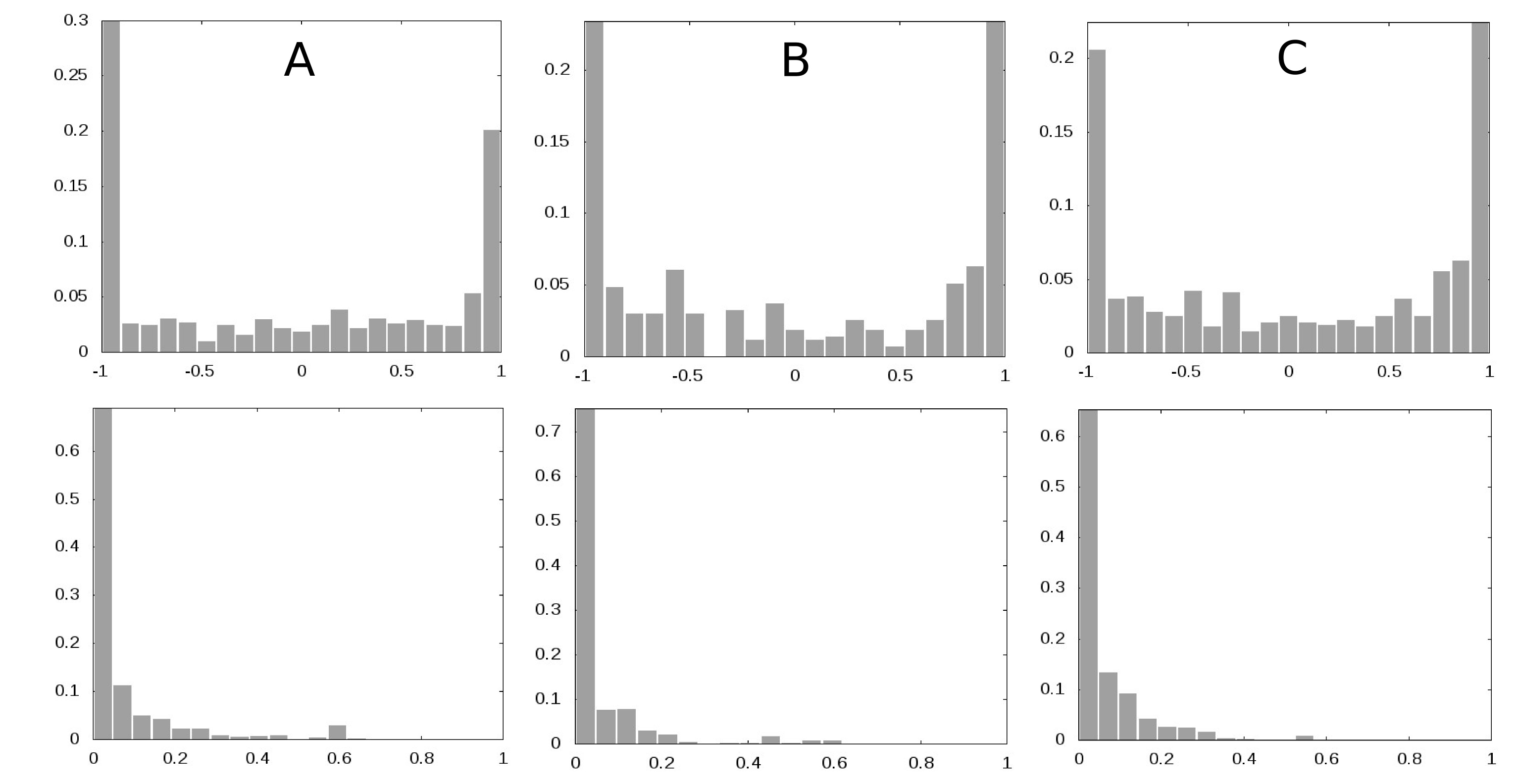}
  \caption{Empirical distributions of CD-s (first row) and overlaps (second row) for the excitatory (column A), inhibitory (column B) and neutral (column C) edges of the neural signal transduction network.}
  \label{fig:stn_edges}
\end{figure}

Empirical distributions of CD-s and relative overlaps over the excitatory, inhibitory and neutral edge classes in the signal transduction network are shown in Figure \ref{fig:stn_edges}. 

\section*{Acknowledgement}\label{sec:ack} Authors are grateful to Tam\'as Nepusz, L\'aszl\'o Ketskem\'ety, L\'aszl\'o Zal\'anyi, Bal\'azs Ujfalussy, Gerg\H{o} Orb\'an and Zolt\'an Somogyv\'ari for useful discussions.

\end{document}